\PassOptionsToPackage{unicode}{hyperref}
\PassOptionsToPackage{hyphens}{url}
\documentclass[
]{article}
\usepackage{amsmath,amssymb}
\usepackage{iftex}
\ifPDFTeX
  \usepackage[T1]{fontenc}
  \usepackage[utf8]{inputenc}
  \usepackage{textcomp} 
\else 
  \usepackage{unicode-math} 
  \defaultfontfeatures{Scale=MatchLowercase}
  \defaultfontfeatures[\rmfamily]{Ligatures=TeX,Scale=1}
\fi
\usepackage{lmodern}
\ifPDFTeX\else
\fi
\IfFileExists{upquote.sty}{\usepackage{upquote}}{}
\IfFileExists{microtype.sty}{
  \usepackage[]{microtype}
  \UseMicrotypeSet[protrusion]{basicmath} 
}{}
\makeatletter
\@ifundefined{KOMAClassName}{
  \IfFileExists{parskip.sty}{%
    \usepackage{parskip}
  }{
    \setlength{\parindent}{0pt}
    \setlength{\parskip}{6pt plus 2pt minus 1pt}}
}{
  \KOMAoptions{parskip=half}}
\makeatother
\usepackage{xcolor}
\usepackage[margin=1in]{geometry}
\usepackage{color}
\usepackage{fancyvrb}

\DefineVerbatimEnvironment{Highlighting}{Verbatim}{commandchars=\\\{\}}
\newenvironment{Shaded}{}{}

\newcommand{\CommentTok}[1]{\textcolor[rgb]{0.38,0.63,0.69}{\textit{#1}}}

\newcommand{\ControlFlowTok}[1]{\textcolor[rgb]{0.00,0.44,0.13}{\textbf{#1}}}

\newcommand{\DecValTok}[1]{\textcolor[rgb]{0.25,0.63,0.44}{#1}}

\newcommand{\KeywordTok}[1]{\textcolor[rgb]{0.00,0.44,0.13}{\textbf{#1}}}
\newcommand{\NormalTok}[1]{#1}
\newcommand{\OperatorTok}[1]{\textcolor[rgb]{0.40,0.40,0.40}{#1}}

\newcommand{\StringTok}[1]{\textcolor[rgb]{0.25,0.44,0.63}{#1}}

\providecommand{\tightlist}{%
  \setlength{\itemsep}{0pt}\setlength{\parskip}{0pt}}
\ifLuaTeX
  \usepackage{selnolig}  
\fi
\IfFileExists{bookmark.sty}{\usepackage{bookmark}}{\usepackage{hyperref}}
\IfFileExists{xurl.sty}{\usepackage{xurl}}{} 
\hypersetup{
  pdftitle={C-lisp and Flexible Macro Programming with S-expressions},
  pdfauthor={Vedanth Padmaraman, Sasank Chilamkurthy},
  hidelinks,
  pdfcreator={LaTeX via pandoc}}

\title{C-lisp and Flexible Macro Programming with S-expressions}
\author{Vedanth Padmaraman, Sasank Chilamkurthy}
\date{}

\begin{document}
\maketitle
\begin{abstract}
Llama.lisp is a compiler framework intended to target offload processor
backends such as GPUs, using intermediate representation languages (IRs)
that are device-agnostic. The Llama.lisp IRs are formulated as
S-expressions. This makes them easy to generate using higher level
programming languages, which is one of the primary goals for Llama.lisp.
The highest IR layer currently implemented in Llama.lisp is C-Lisp. In
this paper, we describe the macro system developed for the Llama.lisp
compiler framework. We show how we implemented FFI bindings as an
example of this system.
\end{abstract}

Compilers are workhorses of performance behind all AI algorithms. Making
algorithms work effectively on GPUs is especially hard -- called kernel
programming. The compiler ecosystem around GPUs is especially
fragmented. They are supposed to allow for performance portability
between different hardware architecture. Unfortunately, this is usually
not the case.

We are designing a compiler framework called llama.lisp {[}1{]} to solve
this problem. As suggested by the name, the framework is highly inspired
by Lisp and its syntax, S-expressions. A multi layered approach is
adopted to tame the complexity of writing such a compiler framework. We
implement C-lisp as one such layer. We show how lisp syntax has allowed
for unique meta programming capabilities while being simple both to
understand and implement.

\hypertarget{c-lisp-structured-llvm-ir}{%
\subsection{1. C-Lisp: Structured LLVM
IR}\label{c-lisp-structured-llvm-ir}}

C-Lisp serves as a structured programming {[}2{]} interface to the LLVM
{[}3{]} instruction set, with semantics modelled after the C language
{[}4{]}. The S-expression syntax forms the base of the C-Lisp syntax. An
S-expression can be either a token or a list, the elements of which are
also S-expressions. The first element of a list usually specifies an
action (in which case it is a token), and the remainder of the elements
specify the arguments to that action. By a slight extension of logic,
S-expressions can also be viewed as trees: a list represents an internal
node, the first element of the list the node type, and the remainder of
the elements the node's children. For example, consider the following
variable declaration in C:

\begin{Shaded}
\begin{Highlighting}[]
\NormalTok{int var;}
\end{Highlighting}
\end{Shaded}

The root node of the abstract syntax tree (AST) for this statement is a
\emph{declaration} node; the children of the root node are the type
\texttt{int} and the variable reference \texttt{var}. One could
represent this AST using S-expressions like so:

\begin{verbatim}
(declare var int)
\end{verbatim}

And it so happens that this is the exact syntax for variable
declarations in C-Lisp.

Most expression opcodes in C-Lisp (i.e.~directives that specify some
computation) exhibit a close correspondence to instruction opcodes in
the LLVM IR, in that they perform the same operations and take the same
kinds of arguments. For example, the LLVM IR implements the
\texttt{fadd} opcode for integer addition, with the syntax

\begin{Shaded}
\begin{Highlighting}[]
\NormalTok{\textless{}result\textgreater{} = fadd [fast{-}math flags]* \textless{}ty\textgreater{} \textless{}op1\textgreater{}, \textless{}op2\textgreater{}}
\end{Highlighting}
\end{Shaded}

C-Lisp exposes a single form of this instruction, consisting of the
compulsory operands, through its \texttt{fadd} expression opcode:

\begin{Shaded}
\begin{Highlighting}[]
\NormalTok{(fadd \textless{}op1\textgreater{} \textless{}op2\textgreater{})}
\end{Highlighting}
\end{Shaded}

Owing to the adoption of C semantics, it can be noted that the result is
not specified in the \texttt{fadd} expression; the \texttt{set} opcode
fulfills that purpose, and can be used with the \texttt{fadd} expression
as an operand. Additionally, the type is inferred, not explicitly
stated.

As an illustration of C-Lisp, consider the following C function to add
the product of two numbers to the contents of a pointer. The function
returns nothing, takes one pointer to a 64-bit integer and two 32-bit
integers as arguments (the bit widths are platform-specific, but we
shall assume these).

\begin{Shaded}
\begin{Highlighting}[]
\NormalTok{void muladd (long int * res, int a, int b) \{}
\NormalTok{    int mul\_res = a * b;}
\NormalTok{    *res = *res + mul\_res;}
\NormalTok{\}}
\end{Highlighting}
\end{Shaded}

An equivalent C-Lisp implementation would be:

\begin{verbatim}
(define ((muladd void) (res (ptr int64)) (a int) (b int))
    (declare mul_res int)
    (set mul_res (mul a b))
    (store res (add (load res) (sext mul_res int64))))
\end{verbatim}

On the face of it, there is a world of difference between the two
versions. However, on closer observation, the C-Lisp version closely
resembles the AST of the C version. Consider the assignment of
\texttt{mul\_res} in C: it is an assignment expression with
\texttt{mul\_res} as its first operand and \texttt{a\ *\ b} as its
second. Further recursing into the second operand, it is a
multiplication expression with \texttt{a} and \texttt{b} as operands.
The C-Lisp version reflects this structure accurately, with \texttt{set}
denoting an assignment and \texttt{mul} denoting a multiplication.

As a result, both implementations have similar semantics, and the
executables produced from both perform equally well. However, the
adoption of S-expressions makes it much more conducive to generate and
programmatically interact with the C-Lisp version.

One main point of difference between semantics of two versions is the
use of implicit casting. The C version adds \texttt{mul\_res}, a 32-bit
integer, to the contents of \texttt{res}, a 64-bit integer. This works
because a compliant C compiler will insert an implicit cast from a 32-
to a 64-bit integer, and thus behave as if the source program had stated

\begin{Shaded}
\begin{Highlighting}[]
\NormalTok{*res = *res + (long int) mul\_res;}
\end{Highlighting}
\end{Shaded}

C-Lisp, on the other hand, employs no implicit action whatsoever. The
programmer is forced to explicitly cast \texttt{mul\_res} to a 64-bit
integer. This helps keep the C-Lisp language's implementation concise
and simple. Additionally, the absence of implicit actions simplifies the
analysis of these programs.

To ease the process of C-Lisp code generation, the JavaScript Object
Notation (JSON) is used as an exchange format for C-Lisp. JSON has
support for lists as well as the basic token types (integers,
floating-point numbers and so on), which makes it an ideal choice for
serializing S-expressions. Additionally, JSON enjoys support in most
mature programming languages. The transformer from S-expression to JSON
is written in Guile Scheme, and as such uses most of Scheme's
conventions for capturing constructs such as \texttt{unquote}.

\hypertarget{a-macro-preprocessor}{%
\subsection{2. A Macro Preprocessor}\label{a-macro-preprocessor}}

C-Lisp is intended to be minimal; most computation can be expressed in
C-Lisp with reasonably simple code, and there is seldom more than one
way to do so. This necessitates a strong macro system: one that enables
extensions of C-Lisp, reducing the need for feature additions to the
language. Prelisp aims to fulfill this need, borrowing from the
multistage programming {[}5{]} paradigm.

Prelisp uses Python as the macro language, although any modern
general-purpose language could have been used. On the face of it, using
a third-party language for the preprocessor can make for rather
complicated macro definitions; however, owing to the adoption of the
S-expression syntactical form, the process of C-Lisp code generation is
greatly simplified. Thus, Python's own \texttt{list} data structure make
it feasible to programmatically emit C-Lisp code. Additionally, Python
makes for a good choice because it involves a minimal learning curve,
and it leaves a powerful standard library and programming environment at
the macro programmer's disposal.

The Prelisp preprocessor takes the input program as a JSON object.
Portions of this object are recognized as macro expressions, evaluated
using macro definitions from a supplied Python module (the ``macro
module'' henceforth), and replaced to produce the result. A macro is
expected to be defined in the global scope of the macro module, and is
either referenced directly, like a variable, or called, like a function.
In both cases, the macro evaluates to a Python object which is
substituted in place of the macro expression and eventually serialized
back into JSON along with the rest of the program. Macro expressions in
the source program are denoted using either the \texttt{unquote} or the
\texttt{unquote-splicing} constructs {[}6{]}, borrowed from the Lisp
family.

\hypertarget{variable-substitution}{%
\subsubsection{2.1. Variable substitution}\label{variable-substitution}}

\texttt{unquote} can be used to substitute a single expression. The
following expression

\begin{verbatim}
; In the source program
(eq (call getchar) ,EOF)
\end{verbatim}

is equivalent to the S-expression

\begin{verbatim}
(eq (call getchar) (unquote EOF))
\end{verbatim}

and thus is represented in JSON as

\begin{Shaded}
\begin{Highlighting}[]
\NormalTok{[}\StringTok{"eq"}\OperatorTok{,}\NormalTok{ [}\StringTok{"call"}\OperatorTok{,} \StringTok{"getchar"}\NormalTok{]}\OperatorTok{,}\NormalTok{ [}\StringTok{"unquote"}\OperatorTok{,} \StringTok{"EOF"}\NormalTok{]]}
\end{Highlighting}
\end{Shaded}

Given this macro expression, Prelisp recognizes \texttt{EOF} as the
unquoted expression and looks for an object named \texttt{EOF} in the
global scope of the macro module. With the following definition in the
macro module

\begin{Shaded}
\begin{Highlighting}[]
\CommentTok{\# In the macro module}
\NormalTok{EOF }\OperatorTok{=}\NormalTok{ [}\StringTok{"trunc"}\NormalTok{, }\OperatorTok{{-}}\DecValTok{1}\NormalTok{, }\StringTok{"int8"}\NormalTok{]}
\end{Highlighting}
\end{Shaded}

the macro expression evaluates to

\begin{Shaded}
\begin{Highlighting}[]
\NormalTok{[}\StringTok{"eq"}\OperatorTok{,}\NormalTok{ [}\StringTok{"call"}\OperatorTok{,} \StringTok{"getchar"}\NormalTok{]}\OperatorTok{,}\NormalTok{  [}\StringTok{"trunc"}\OperatorTok{,} \OperatorTok{{-}}\DecValTok{1}\OperatorTok{,} \StringTok{"int8"}\NormalTok{]]}
\end{Highlighting}
\end{Shaded}

and when converted back to S-expression form yields

\begin{verbatim}
(eq (call getchar) (trunc -1 int8))
\end{verbatim}

\hypertarget{parametric-macros}{%
\subsubsection{2.2. Parametric macros}\label{parametric-macros}}

Consider a function call-like macro expression:

\begin{verbatim}
; In the source program
,(incr var 45)
\end{verbatim}

with the equivalent JSON form:

\begin{Shaded}
\begin{Highlighting}[]
\NormalTok{[}\StringTok{"unquote"}\OperatorTok{,}\NormalTok{ [}\StringTok{"incr"}\OperatorTok{,} \StringTok{"var"}\OperatorTok{,} \DecValTok{45}\NormalTok{]]}
\end{Highlighting}
\end{Shaded}

and a corresponding definition in the macro module:

\begin{Shaded}
\begin{Highlighting}[]
\CommentTok{\# In the macro module}
\KeywordTok{def}\NormalTok{ incr (name, amt)}
    \CommentTok{"""(incr name, amt) {-}\textgreater{} (set name (add name amt))"""}
    \ControlFlowTok{return}\NormalTok{ [}\StringTok{"set"}\NormalTok{, name, [}\StringTok{"add"}\NormalTok{, name, amt]]}
\end{Highlighting}
\end{Shaded}

Since the expression after \texttt{unquote} is a list, Prelisp infers
\texttt{incr} to be the name of a callable in the macro module. The
macro is evaluated by calling \texttt{incr} with arguments
\texttt{"var"} and \texttt{45}, and the resulting macro substitution's
JSON form looks like this:

\begin{Shaded}
\begin{Highlighting}[]
\NormalTok{[}\StringTok{"set"}\OperatorTok{,} \StringTok{"var"}\OperatorTok{,}\NormalTok{ [}\StringTok{"add"}\OperatorTok{,} \StringTok{"var"}\OperatorTok{,} \DecValTok{45}\NormalTok{]]}
\end{Highlighting}
\end{Shaded}

When converted back to the S-expression form:

\begin{verbatim}
(set var (add var 45))
\end{verbatim}

\hypertarget{splicing-macros}{%
\subsubsection{2.3. Splicing macros}\label{splicing-macros}}

\texttt{unquote-splicing} can be used to substitute multiple expressions
in place of a single macro expression. An expression of the form

\begin{verbatim}
; In the source program
,@(declare_multiple (ch i) int)
\end{verbatim}

is represented in JSON as

\begin{Shaded}
\begin{Highlighting}[]
\NormalTok{[}\StringTok{"unquote{-}splicing"}\OperatorTok{,}\NormalTok{ [}\StringTok{"declare\_multiple"}\OperatorTok{,}\NormalTok{ [}\StringTok{"ch"}\OperatorTok{,} \StringTok{"i"}\NormalTok{]}\OperatorTok{,} \StringTok{"int"}\NormalTok{]]}
\end{Highlighting}
\end{Shaded}

Given the following macro definition,

\begin{Shaded}
\begin{Highlighting}[]
\CommentTok{\# In the macro module}
\KeywordTok{def}\NormalTok{ declare\_multiple(names, typ):}
\NormalTok{    decls }\OperatorTok{=}\NormalTok{ []}
    \ControlFlowTok{for}\NormalTok{ name }\KeywordTok{in}\NormalTok{ names:}
\NormalTok{        decls.append([}\StringTok{"declare"}\NormalTok{, name, typ])}
    \ControlFlowTok{return}\NormalTok{ decls}
\end{Highlighting}
\end{Shaded}

The macro expression is replaced with

\begin{Shaded}
\begin{Highlighting}[]
\NormalTok{[}\StringTok{"declare"}\OperatorTok{,} \StringTok{"ch"}\OperatorTok{,} \StringTok{"int"}\NormalTok{]}
\NormalTok{[}\StringTok{"declare"}\OperatorTok{,} \StringTok{"i"}\OperatorTok{,} \StringTok{"int"}\NormalTok{]}
\end{Highlighting}
\end{Shaded}

Thus, in S-expression, this looks like

\begin{verbatim}
(declare ch int)
(declare i int)
\end{verbatim}

Note that if \texttt{unquote} (i.e.~\texttt{,} instead of \texttt{,@})
was used, both of the declare statements would be nested under a list,
like so:

\begin{verbatim}
((declare ch int)
 (declare i int))
\end{verbatim}

Note that the return values of \texttt{incr} and
\texttt{declare\_multiple} are entirely composed of native Python data
structures, and the literal expressions used to construct the return
values closely resemble the actual S-expressions that are emitted. This
highlights the ease of C-Lisp code generation.

\hypertarget{example-building-an-ffi-system-using-prelisp}{%
\subsection{3. Example: Building an FFI System using
Prelisp}\label{example-building-an-ffi-system-using-prelisp}}

C-Lisp is compatible with C at the ABI level. This means that libraries
that can be used with C code can also be used with C-Lisp in a similar
fashion. In C, using an external library typically involves placing
forward definitions for the library's contents in the source program,
and linking to the library's object file; the same holds for C-Lisp too.

Libraries are typically distributed along with header files containing
forward declarations for their contents. C's \texttt{\#include}
preprocessor directive is typically the mechanism by which the forward
declarations from these header files are brought into the source of a
program that uses the library. Since C-Lisp uses C's data types, it is
feasible to generate forward declarations in C-Lisp from forward
declarations in C; consequently, a library's C header files can be used
to generate C-Lisp bindings to the library.

Prelisp makes it possible to implement a solution for binding generation
entirely in Python and expose it as a macro for use in a C-Lisp program.
Such a solution is under active development, and is already in use by a
test program that launches accelerated vector addition on an NVIDIA GPU
using the CUDA driver API.

Parsing C is a relatively complex task, partly due to C's complicated
syntax, and partly due to the presence of constructs in the C language
that are outside the scope of C-Lisp --- \texttt{typedef},
\texttt{enum}, and so on. For these reasons, the actual parsing of C
code is offloaded to the Clang frontend. Clang is used to produce two
artifacts from a C header: the LLVM IR module and the AST in Clang's own
JSON schema. The LLVM IR is then parsed using Numba's {[}7{]} LLVMLite
binding layer to yield function declarations and struct type definitions
(collectively referred to as ``signatures'' henceforth), while type
aliases (\texttt{typedef}s) are scraped from the JSON AST.

The binding generation process works on this premise. A Python module
orchestrates the processes of running the Clang executable, saving its
outputs, and processing the LLVM IR and the AST to yield declarations in
C-Lisp. The process is as follows:

\begin{itemize}
\tightlist
\item
  Take input for desired headers, functions, structs and typedefs
\item
  Generate a C program that

  \begin{itemize}
  \tightlist
  \item
    includes the desired header files
  \item
    uses each of the desired functions and structs
  \end{itemize}
\item
  Compile the generated C program, saving its JSON AST and LLVM IR
\item
  Parse the IR to extract function and struct type signatures
\item
  Parse the JSON AST to extract typedef type aliases and function
  parameter names
\end{itemize}

This same module, when used as a Prelisp macro module, serves as a
convenient means of using definitions from external libraries. At
present, its usage on the CUDA driver API is a single macro call:

\begin{verbatim}
    ,@(include
        (/usr/local/cuda/include/cuda.h) ; Headers
        (cuInit
         cuDeviceGetCount
         cuDeviceGet
         cuCtxCreate_v2
         cuModuleLoadDataEx
         cuModuleGetFunction
         cuMemAlloc_v2
         cuMemcpyHtoD_v2
         cuLaunchKernel
         cuCtxSynchronize
         cuMemcpyDtoH_v2
         cuMemFree_v2
         cuModuleUnload
         cuCtxDestroy_v2) ; Functions
        () ; Structs
        (CUcontext CUmodule CUfunction CUstream CUdevice)) ; Typedefs
\end{verbatim}

And this allows access to the CUDA driver API through rather familiar
names:

\begin{verbatim}
(declare module ,CUmodule)
(declare kernel_func ,CUfunction)
; ...
(call cuModuleGetFunction (ptr-to kernel_func) module "kernel")
\end{verbatim}

For reference, the equivalent C version would look like this:

\begin{Shaded}
\begin{Highlighting}[]
\NormalTok{\#include \textless{}cuda.h\textgreater{}}

\NormalTok{CUmodule module;}
\NormalTok{CUfunction kernel\_func;}
\NormalTok{// ...}
\NormalTok{cuModuleGetFunction(\&kernel\_func, module, "kernel");}
\end{Highlighting}
\end{Shaded}

\hypertarget{conclusion}{%
\subsection{4. Conclusion}\label{conclusion}}

The implementation of the Prelisp preprocessor system is a rather
straightforward extension of the ideas it builds on, such as
S-expression IRs and substitution using \texttt{unquote}. However, the
combination of these ideas results in a powerful framework that made it
possible to achieve on-the-fly bindings generation and inclusion with a
few lines of Python code and minimal external dependencies.

\hypertarget{references}{%
\subsection{5. References}\label{references}}

\begin{enumerate}
\def\labelenumi{\arabic{enumi}.}
\tightlist
\item
  The Llama.lisp Compiler Framework.
  https://github.com/chsasank/llama.lisp
\item
  Dijkstra, Edsger W. ``Letters to the editor: go to statement
  considered harmful.'' \emph{Communications of the ACM} 11.3 (1968):
  147-148.
\item
  Lattner, Chris, and Vikram Adve. ``LLVM: A compilation framework for
  lifelong program analysis \& transformation.'' \emph{International
  symposium on code generation and optimization, 2004. CGO 2004.}. IEEE,
  2004.
\item
  Kernighan, Brian W., and Dennis M. Ritchie. \emph{The C programming
  language}. prentice-Hall, 1988.
\item
  Taha, Walid. ``A gentle introduction to multi-stage programming.''
  \emph{Domain-Specific Program Generation: International Seminar,
  Dagstuhl Castle, Germany, March 23-28, 2003. Revised Papers}. Berlin,
  Heidelberg: Springer Berlin Heidelberg, 2004.
\item
  Bawden, Alan. ``Quasiquotation in Lisp.'' \emph{PEPM}. 1999.
\item
  Lam, Siu Kwan, Antoine Pitrou, and Stanley Seibert. ``Numba: A
  llvm-based python jit compiler.'' \emph{Proceedings of the Second
  Workshop on the LLVM Compiler Infrastructure in HPC}. 2015.
\end{enumerate}

\end{document}